\documentstyle[art10]{article}

\begin{document}
\def\rf#1{(\ref{eq:#1})}
\def\lab#1{\label{eq:#1}}
\def\nonu{\nonumber}
\def\br{\begin{eqnarray}}
\def\er{\end{eqnarray}}
\def\be{\begin{equation}}
\def\ee{\end{equation}}
\def\eq{\!\!\!\! &=& \!\!\!\! }
\def\foot#1{\footnotemark\footnotetext{#1}}
\def\lb{\lbrack}
\def\rb{\rbrack}
\def\llangle{\left\langle}
\def\rrangle{\right\rangle}
\def\blangle{\Bigl\langle}
\def\brangle{\Bigr\rangle}
\def\llb{\left\lbrack}
\def\rrb{\right\rbrack}
\def\Blb{\Bigl\lbrack}
\def\Brb{\Bigr\rbrack}
\def\lcurl{\left\{}
\def\rcurl{\right\}}
\def\({\left(}
\def\){\right)}
\def\v{\vert}                     
\def\bv{\bigm\vert}               
\def\Bgv{\;\Bigg\vert}            
\def\bgv{\bigg\vert}              
\def\lskip{\vskip\baselineskip\vskip-\parskip\noindent}
\def\mskp{\par\vskip 0.3cm \par\noindent}
\def\sskp{\par\vskip 0.15cm \par\noindent}
\def\bc{\begin{center}}
\def\ec{\end{center}}
\def\Lbf#1{{\Large {\bf {#1}}}}
\def\lbf#1{{\large {\bf {#1}}}}


\def\tr{\mathop{\rm tr}}                  
\def\Tr{\mathop{\rm Tr}}                  
\newcommand\partder[2]{{{\partial {#1}}\over{\partial {#2}}}}
\newcommand\partderd[2]{{{\partial^2 {#1}}\over{{\partial {#2}}^2}}}
\newcommand\partderh[3]{{{\partial^{#3} {#1}}\over{{\partial {#2}}^{#3} }}}
\newcommand\partderm[3]{{{\partial^2 {#1}}\over{\partial {#2} \partial {#3} }}}
\newcommand\partderM[6]{{{\partial^{#2} {#1}}\over{{\partial {#3}}^{#4}
{\partial {#5}}^{#6} }}}          
\newcommand\funcder[2]{{{\delta {#1}}\over{\delta {#2}}}}
\newcommand\Bil[2]{\Bigl\langle {#1} \Bigg\vert {#2} \Bigr\rangle}  
\newcommand\bil[2]{\left\langle {#1} \bigg\vert {#2} \right\rangle} 
\newcommand\me[2]{\left\langle {#1}\right|\left. {#2} \right\rangle} 

\newcommand\sbr[2]{\left\lbrack\,{#1}\, ,\,{#2}\,\right\rbrack} 
\newcommand\Sbr[2]{\Bigl\lbrack\,{#1}\, ,\,{#2}\,\Bigr\rbrack} 
\newcommand\Gbr[2]{\Bigl\lbrack\,{#1}\, ,\,{#2}\,\Bigr\} } 
\newcommand\pbr[2]{\{\,{#1}\, ,\,{#2}\,\}}       
\newcommand\Pbr[2]{\Bigl\{ \,{#1}\, ,\,{#2}\,\Bigr\}}  
\newcommand\pbbr[2]{\lcurl\,{#1}\, ,\,{#2}\,\rcurl}  


\def\a{\alpha}
\def\b{\beta}
\def\c{\chi}
\def\d{\delta}
\def\D{\Delta}
\def\eps{\epsilon}
\def\vareps{\varepsilon}
\def\g{\gamma}
\def\G{\Gamma}
\def\grad{\nabla}
\def\h{{1\over 2}}
\def\k{\kappa}
\def\l{\lambda}
\def\L{\Lambda}
\def\m{\mu}
\def\n{\nu}
\def\o{\over}
\def\om{\omega}
\def\O{\Omega}
\def\p{\phi}
\def\P{\Phi}
\def\pa{\partial}
\def\tpa{{\tilde \partial}}
\def\bpa{{\bar \partial}}
\def\pr{\prime}
\def\ra{\rightarrow}
\def\lra{\longrightarrow}
\def\s{\sigma}
\def\S{\Sigma}
\def\t{\tau}
\def\th{\theta}
\def\Th{\Theta}
\def\z{\zeta}
\def\ti{\tilde}
\def\wti{\widetilde}
\newcommand\sumi[1]{\sum_{#1}^{\infty}}   
\newcommand\twomat[4]{\left(\begin{array}{cc}  
{#1} & {#2} \\ {#3} & {#4} \end{array} \right)}
\newcommand\threemat[9]{\left(\begin{array}{ccc}  
{#1} & {#2} & {#3}\\ {#4} & {#5} & {#6}\\
{#7} & {#8} & {#9} \end{array} \right)}
\newcommand\BDet[5]{\det_{{#1}}\left\Vert\begin{array}{cc}  
{#2} & {#3} \\ {#4} & {#5} \end{array} \right\Vert}   
\newcommand\Det[2]{\det_{{#1}} \left\Vert {#2} \right\Vert}
\newcommand\twocol[2]{\left(\begin{array}{cc}  
{#1} \\ {#2} \end{array} \right)}


\def\cA{{\cal A}}
\def\cB{{\cal B}}
\def\cC{{\cal C}}
\def\cD{{\cal D}}
\def\cE{{\cal E}}
\def\cF{{\cal F}}
\def\cG{{\cal G}}
\def\cH{{\cal H}}
\def\cI{{\cal I}}
\def\cJ{{\cal J}}
\def\cK{{\cal K}}
\def\cL{{\cal L}}
\def\cM{{\cal M}}
\def\cN{{\cal N}}
\def\cO{{\cal O}}
\def\cP{{\cal P}}
\def\cQ{{\cal Q}}
\def\cR{{\cal R}}
\def\cS{{\cal S}}
\def\cT{{\cal T}}
\def\cU{{\cal U}}
\def\cV{{\cal V}}
\def\cX{{\cal X}}
\def\cW{{\cal W}}
\def\cY{{\cal Y}}
\def\cZ{{\cal Z}}


\def\mark{\noindent{\bf Remark.}\quad}
\def\prop{\noindent{\bf Proposition.}\quad}
\def\exam{\noindent{\bf Example.}\quad}

\newtheorem{definition}{Definition}
\newtheorem{proposition}{Proposition}
\newtheorem{theorem}{Theorem}
\newtheorem{lemma}{Lemma}
\newtheorem{corollary}{Corollary}

\newcommand\Back{{B\"{a}cklund}~}
\newcommand\DB{{Darboux-B\"{a}cklund}~}
\def\cKPrm{${\sf cKP}_{r,m}$~}
\def\vp{{\varphi}}
\def\bt{{\bar t}}
\def\pai{\partial^{-1}}
\def\bD{{\bar D}}
\def\bpai{{\bar \partial}^{-1}}
\def\bcL{{\bar {\cal L}}}
\def\bP{{\bar \Phi}}
\def\bPsi{{\bar \Psi}}
\def\Dth{\cD_\theta}
\def\sRes{{\cal R}es}
\def\SKPrm{${\sl SKP}_{{r\o 2},{m\o 2}}$~}
\def\SKPhh{${\sl SKP}_{\h,\h}$~}

\newcommand{\nit}{\noindent}
\newcommand{\ct}[1]{\cite{#1}}
\newcommand{\bi}[1]{\bibitem{#1}}
\newcommand\PRL[3]{{\sl Phys. Rev. Lett.} {\bf#1} (#2) #3}
\newcommand\NPB[3]{{\sl Nucl. Phys.} {\bf B#1} (#2) #3}
\newcommand\NPBFS[4]{{\sl Nucl. Phys.} {\bf B#2} [FS#1] (#3) #4}
\newcommand\CMP[3]{{\sl Commun. Math. Phys.} {\bf #1} (#2) #3}
\newcommand\PRD[3]{{\sl Phys. Rev.} {\bf D#1} (#2) #3}
\newcommand\PLA[3]{{\sl Phys. Lett.} {\bf #1A} (#2) #3}
\newcommand\PLB[3]{{\sl Phys. Lett.} {\bf #1B} (#2) #3}
\newcommand\JMP[3]{{\sl J. Math. Phys.} {\bf #1} (#2) #3}
\newcommand\PTP[3]{{\sl Prog. Theor. Phys.} {\bf #1} (#2) #3}
\newcommand\SPTP[3]{{\sl Suppl. Prog. Theor. Phys.} {\bf #1} (#2) #3}
\newcommand\AoP[3]{{\sl Ann. of Phys.} {\bf #1} (#2) #3}
\newcommand\RMP[3]{{\sl Rev. Mod. Phys.} {\bf #1} (#2) #3}
\newcommand\PR[3]{{\sl Phys. Reports} {\bf #1} (#2) #3}
\newcommand\FAP[3]{{\sl Funkt. Anal. Prilozheniya} {\bf #1} (#2) #3}
\newcommand\FAaIA[3]{{\sl Functional Analysis and Its Application} {\bf #1}
(#2) #3}
\def\TAMS#1#2#3{{\sl Trans. Am. Math. Soc.} {\bf #1} (#2) #3}
\def\InvM#1#2#3{{\sl Invent. Math.} {\bf #1} (#2) #3}
\def\AdM#1#2#3{{\sl Advances in Math.} {\bf #1} (#2) #3}
\def\PNAS#1#2#3{{\sl Proc. Natl. Acad. Sci. USA} {\bf #1} (#2) #3}
\newcommand\LMP[3]{{\sl Letters in Math. Phys.} {\bf #1} (#2) #3}
\newcommand\IJMPA[3]{{\sl Int. J. Mod. Phys.} {\bf A#1} (#2) #3}
\newcommand\TMP[3]{{\sl Theor. Mat. Phys.} {\bf #1} (#2) #3}
\newcommand\JPA[3]{{\sl J. Physics} {\bf A#1} (#2) #3}
\newcommand\JSM[3]{{\sl J. Soviet Math.} {\bf #1} (#2) #3}
\newcommand\MPLA[3]{{\sl Mod. Phys. Lett.} {\bf A#1} (#2) #3}
\newcommand\JETP[3]{{\sl Sov. Phys. JETP} {\bf #1} (#2) #3}
\newcommand\JETPL[3]{{\sl  Sov. Phys. JETP Lett.} {\bf #1} (#2) #3}
\newcommand\PHSA[3]{{\sl Physica} {\bf A#1} (#2) #3}
\newcommand\PHSD[3]{{\sl Physica} {\bf D#1} (#2) #3}
\newcommand\JPSJ[3]{{\sl J. Phys. Soc. Jpn.} {\bf #1} (#2) #3}
\newcommand\JGP[3]{{\sl J. Geom. Phys.} {\bf #1} (#2) #3}
  
\setlength{\baselineskip}{2.6ex}

\title{Berezinian Construction of Super-Solitons in
Supersymmetric Constrained KP Hierarchies}

\author{H. Aratyn${}^1$, E. Nissimov${}^{2,3}$ and S. Pacheva${}^{2,3}$ \\
{\em ${}^1$ Department of Physics, University of Illinois at Chicago}\\
{\em 845 W. Taylor St., Chicago, IL 60607-7059, U.S.A.}\\
{\em ${}^2$ Institute of Nuclear Research and Nuclear Energy} \\
{\em Boul. Tsarigradsko Chausee 72, BG-1784 $\;$Sofia, Bulgaria}\\ 
{\em ${}^3$Department of Physics, Ben-Gurion University of the Negev} \\
{\em Box 653, IL-84105 $\;$Beer-Sheva, Israel}}

\maketitle
\begin{abstract}
\setlength{\baselineskip}{2.6ex}   

We consider a broad class \SKPrm of consistently reduced 
Manin-Radul supersymmetric KP hierarchies (MR-SKP) which are supersymmetric
analogs of the ordinary bosonic constrained KP models. Compatibility of these
reductions to \SKPrm with the MR fermionic isospectral flows is achieved via
appropriate modification of the latter preserving their (anti-)commutation
algebra. Unlike the general unconstrained MR-SKP case, \DB transformations 
do preserve the fermionic isospectral flows of \SKPrm . This allows for a 
systematic derivation of explicit Berezinian solutions for the \SKPrm 
super-tau-functions (super-solitons).
\end{abstract}  
\setlength{\baselineskip}{2.6ex}

\section*{Introduction}

Manin-Radul supersymmetric KP (MR-SKP) integrable hierarchy of nonlinear
evolution (``super-soliton'') equations \ct{MR-SKP} and other related
supersymmetric integrable hierarchies 
\ct{Susy-SG,SI-KdV-NLS,SKP-other-1,SKP-other,SKP-2,SKP-red,AR97} attracted
a lot of interest, both from purely mathematical point of view as
supersymmetric generalizations of the inverse scattering method,
bi-Hamiltonian structures, tau-functions and Sato Grassmannian approach, as
well as in the context of theoretical physics due to their relevance in
non-perturbative superstring theory \ct{SI-sstring}.

In the present paper we will be specifically concerned with MR-SKP hierarchy
\ct{MR-SKP}, {\sl i.e.}, possessing $N=1$ supersymmetry and being defined in 
terms of {\em fermionic} (Grassmann-odd) pseudo-differential Lax operator.
In ref.\ct{match} we have already started a systematic study of MR-SKP
hierarchy with particular attention being paid to the proper treatment of the 
fermionic MR isospectral flows, which was lacking in the previous studies on 
the subject. In \ct{match} we introduced an infinite algebra of commuting
additional (``ghost'') symmetries of MR-SKP hierarchy which were used to
construct systematic reductions to a broad class of constrained 
supersymmetric KP hierarchies denoted as \SKPrm (see Eq.\rf{SKP-r-m} below;
we will keep in the sequel the name MR-SKP to explicitly denote the full
unconstrained hierarchy).
The constrained \SKPrm hierarchies possess correct evolution under the bosonic 
(Grassmann-even) isospectral flows. However, it turns out that the reductions 
from MR-SKP to \SKPrm hierarchies are {\em incompatible} with the
original MR-SKP fermionic (Grassmann-odd) isospectral flows. 
In \ct{match} we provided a solution to this problem for the simplest case of
constrained \SKPhh hierarchy by appropriately modifying MR-SKP fermionic flows
while preserving their original (anti-)commutation algebra, {\sl i.e.}, 
preserving the integrability of the constrained \SKPhh system. One of the
results of the present paper is the extention of this construction to all 
\SKPrm hierarchies.

Our next result concerns the construcion of \DB (DB) 
transformations preserving both types (even and odd) of the isospectral flows. 
As already pointed out in \ct{match}, DB transformations are {\em always
incompatible} with the fermionic flows in the original unconstrained MR-SKP
hierarchy. However, for constrained \SKPrm hierarchies the compatibility of
DB transformations is here achieved thanks to the above mentioned modification
of the original MR-SKP fermionic flows.

Furthermore, we provide explicit expressions for the super-tau function and 
the super-eigenfunctions on DB-orbits of iterations of the DB
transformations for arbitrary constrained \SKPrm hierarchies, which are
given in terms of Wronskian-like Berezinians. These Berezinian solutions 
constitute supersymmetric generalizations of the (multi-)so\-li\-ton 
solutions in ordinary bosonic KP hierarchies.

\section*{Background on Manin-Radul Supersymmetric KP Hierarchy}

MR-SKP hierarchy is defined through the {\em fermionic} $N\! =\! 1$ 
super-pseudo-differential Lax operator $\cL$ :
\be
\cL = \cD + f_0 + \sum_{j=1}^\infty b_j \pa^{-j}\cD + 
\sum_{j=1}^\infty f_j \pa^{-j}
\lab{super-Lax}
\ee
where the coefficients $b_j ,f_j$ are bosonic and fermionic superfield
functions, respectively. We shall use throughout this paper
the super-pseudo-differential
calculus \ct{MR-SKP} with the following notations: $\pa$ and
$\cD = \partder{}{\th} + \th \pa$ denote operators, whereas the symbols
$\pa_x$ and $\Dth$ will indicate application of the corresponding operators on
superfield functions. As usual, $(x,\th )$ denote $N\! =\! 1$ superspace 
coordinates.
For any super-pseudo-differential operator $\cA = \sum_j a_{j/2} \cD^j$
the subscripts $(\pm )$ denote its purely differential part
($\cA_{+} = \sum_{j \geq 0} a_{j/2} \cD^j$) or its purely
pseudo-differential part ($\cA_{-} = \sum_{j \geq 1} a_{-j/2} \cD^{-j}$),
respectively. For any $\cA$ the super-residuum is defined as 
$\sRes \cA = a_{-\h}$. 

The Lax evolution Eqs. for MR-SKP read \ct{MR-SKP} :
\br
\partder{}{t_l} \cL &=& -\Sbr{\cL^{2l}_{-}}{\cL} = \Sbr{\cL^{2l}_{+}}{\cL}
\lab{super-Lax-even} \\
D_n \cL &=& -\pbbr{\cL^{2n-1}_{-}}{\cL} =
\pbbr{\cL^{2n-1}_{+}}{\cL} - 2\cL^{2n} 
\lab{super-Lax-odd} 
\er
with the short-hand notations:
\br
D_n = \partder{}{\th_n} - \sum_{k=1}^\infty \th_k \partder{}{t_{n+k-1}}
\quad ,\quad
\pbbr{D_k}{D_l} = - 2 \partder{}{t_{k+l-1}}
\lab{MR-D-n} \\
(t,\th ) \equiv \( t_1 \equiv x ,t_2, \ldots ; \th, \th_1 ,\th_2 ,\ldots \)
\lab{t-th-short}
\er

An important r\^{o}le in the present approach is played by the notion of
(adjoint-) super-eigenfunctions (sEF's) $\P = \P (t,\th)$ and
$\Psi = \Psi (t,\th)$ of MR-SKP hierarchy obeing:
\br 
\partder{}{t_l} \P = \cL^{2l}_{+} (\P) \quad ,\quad
D_n \P = \cL^{2n-1}_{+} (\P) \nonu \\ 
\partder{}{t_l} \Psi = - \(\cL^{2l}\)^{\ast}_{+} (\Psi) \quad ,\quad
D_n \Psi = - \(\cL^{2n-1}\)^{\ast}_{+} (\Psi) 
\lab{super-EF-eqs}
\er
The (adjoint-)super-Baker-Akhiezer functions $\psi_{BA}^{(\ast )}$ of MR-SKP
are particular cases of (adjoint-)sEF's which satisfy the spectral
equations $\(\cL^2\)^{(\ast)} \psi^{(\ast)}_{BA} = \pm \l \psi^{(\ast)}_{BA}$
in addition to \rf{super-EF-eqs}.

Finally, the super-tau-function $\t (t,\th )$ is expressed in terms of the 
super-residues of powers of the super-Lax operator \rf{super-Lax} as follows:
\be
\sRes \cL^{2k} = \partder{}{t_k} \Dth \ln \t \quad ,\quad
\sRes \cL^{2k-1} = D_k \Dth \ln \t
\lab{tau-sres}
\ee

\section*{Constrained Supersymmetric KP Hierarchies}

Let us consider an infinite set $\lcurl\P_{j/2},\Psi_{j/2}\rcurl_{j=0}^\infty$
of pairs of (adjoint-)sEF's of $\cL$ where $j$ indicates their Grassmann
parity (integer indices correspond to bosonic, whereas half-integer indices 
correspond to fermionic parity). It was shown in \ct{match} that
the following infinite set of super-pseudo-differential operators:
\be
\cM_{s/2} = \sum_{k=0}^{s-1} \P_{{s-1-k}\o 2} \cD^{-1} \Psi_{k\o 2}
\quad ,\quad   s=1,2,\ldots
\lab{ghost-s}
\ee
generate an infinite set of flows $\bpa_{s/2}$
($\bpa_{n-\h} \equiv \bD_n \;\; ,\;\; \bpa_k \equiv \partder{}{\bt_k}$) :
\be
\bD_n \cL = \pbbr{\cM_{n-\h}}{\cL} \quad ,\quad
\partder{}{\bt_k} \cL = \Sbr{\cM_k}{\cL}
\lab{ghost-flow-Lax}
\ee
which (anti-)commute with the original isospectral flows $\partder{}{t_l},D_n$
\rf{super-Lax-even}--\rf{super-Lax-odd}, {\sl i.e.},
$\bpa_{s/2}$ define an infinite-dimensional algebra of additional
``ghost'' symmetries of
MR-SKP hierarchy, obeying the (anti-)commutation relations:
\be
\Sbr{\partder{}{\bt_s}}{\partder{}{\bt_k}} = 
\Sbr{\partder{}{\bt_s}}{\bD_n} = 0 \quad ,\quad
\pbbr{\bD_i}{\bD_j} = - 2\partder{}{\bt_{i+j-1}}  
\lab{ghost-alg} 
\ee
The super-``ghost''-symmetry flows and the corresponding generating 
operators $\cM_{s\o 2}$ \rf{ghost-s}--\rf{ghost-flow-Lax} are used to
construct a series of reductions of the MR-SKP hierarchy
\ct{match}. Since the super-``ghost'' flows obey the same algebra 
\rf{ghost-alg} as the original MR-SKP flows \rf{MR-D-n}, one can identify 
an infinite subset of the latter with a corresponding infinite subset of the 
former:
\be
\pa_{\ell {r\o 2}} = - \bpa_{\ell {m\o 2}} \quad ,\;\; \ell =1,2,\ldots \;\; ;
\quad \pa_k \equiv \partder{}{t_k} \; ,\; \pa_{k-\h} \equiv D_k \;\; ;\;\;
\bpa_k \equiv \partder{}{\bt_k} \; ,\; \bpa_{k-\h} \equiv \bD_k
\lab{reduct}
\ee
where $(r,m)$ are some fixed positive integers of {\em equal parity}, 
and retain only these flows as Lax evolution flows (this is a supersymmetric 
extension of the usual reduction procedure in the purely bosonic case 
\ct{Oevel-chengs}). Eqs.\rf{reduct} imply the identification 
$\(\cL^{r\ell}\)_{-} = \cM_{\ell{m\o 2}}$ for any $\ell$. 
Therefore, the pertinent reduced (constrained) MR-SKP hierarchy, denoted 
as \SKPrm , is described by the following constrained super-Lax operator:
\be
\cL_{({r\o 2},{m\o 2})} = \cD^r + \sum_{i=0}^{r-1} v^{(r)}_{i\o 2} \cD^i +
\sum_{j=0}^{m-1} \P_{{m-1-j}\o 2} \cD^{-1} \Psi_{j\o 2}
\lab{SKP-r-m}
\ee
which is the supersymmetric counterpart of the ordinary pseudo-differential 
Lax operator describing the bosonic constrained KP hierarchies \cKPrm (for a
detailed discussion and further references, see \ct{noak}).

Henceforth we will restrict our attention to {\em fermionic} constrained
\SKPrm hierarchies, {\sl i.e.}, \rf{SKP-r-m} with $(r,m)$ being odd integers.

As already pointed out in \ct{match}, the original MR fermionic flows
\rf{super-Lax-odd} are incompatible with the reduction of MR-SKP
\rf{super-Lax} to fermionic constrained \SKPrm hierarchies \rf{SKP-r-m}.
Namely, taking the $(-)$ part of Eqs.\rf{super-Lax-odd} for fermionic
constrained $\cL_{({r\o 2},{m\o 2})}$ (Eq.\rf{SKP-r-m} with $r,m = odd$)
and using a series of simple identities for super-pseudo-differential 
operators \ct{match} we obtain:
\br
\sum_{j=0}^{m-1}\llb \( D_n \P_{{m-1-j}\o 2} - 
\cL^{2n-1}_{+}(\P_{{m-1-j}\o 2})\) \cD^{-1}\Psi_{j\o 2} 
\P_{{m-1-j}\o 2} \cD^{-1} \( D_n \Psi_{j\o 2} + 
\(\cL^{2n-1}\)^\ast_{+}(\Psi_{j\o 2})\)\rrb   \nonu 
\er
\be
= -2 \sum_{j=0}^{m-1} \sum_{k=0}^{2n-1}
\cL^{2n-1-k}(\P_{{m-1-j}\o 2})\cD^{-1}{\cL^k}^\ast (\Psi_{j\o 2})
\lab{naive-D-n}
\ee
which leads to apparent contradiction, since the l.h.s. of \rf{naive-D-n}
vanishes by virtue of Eqs.\rf{super-EF-eqs} for the (adjoint-)sEF's, whereas 
the r.h.s. of \rf{naive-D-n} is manifestly non-zero.

Generalizing the argument given in \ct{match} for 
the simplest \SKPhh case, we arrive at the following: 
\begin{proposition}
There exists the following consistent modification of MR-SKP flows $D_n$
\rf{super-Lax-odd} for constrained \SKPrm hierarchy ($r,m = odd$) :
\br
\cD_k \cL = - \pbbr{\cL^{2k-1}_{-} - X^{(2k-1)}}{\cL} = 
\pbbr{\cL^{2k-1}_{+}}{\cL} + \pbbr{X^{(2k-1)}}{\cL} - 2\cL^{2k}
\lab{odd-flow-new} \\
X^{(2k-1)} \equiv 2 \sum_{j=0}^{m-1} \sum_{l=0}^{k-2} 
\cL^{2(k-l)-3}(\P_{{m-j-1}\o 2} ) \cD^{-1}\(\cL^{2l+1}\)^\ast (\Psi_{j\o 2})
\lab{X-def} \\
\cD_k \P_{j\o 2} = \cL^{2k-1}_{+}(\P_{j\o 2}) - 2 \cL^{2k-1}(\P_{j\o 2})
+ X^{(2k-1)}(\P_{j\o 2})
\lab{P-j-flow-new} \\
\cD_k \Psi_{j\o 2} = - \(\cL^{2k-1}\)^\ast_{+}(\Psi_{j\o 2}) +
2\(\cL^{2k-1}\)^\ast (\Psi_{j\o 2}) - \( X^{(2k-1)}\)^\ast (\Psi_{j\o 2})
\lab{Psi-j-flow-new}
\er
The modified $\cD_k$ flows for \SKPrm obey the same anti-commutation algebra
$\lcurl \cD_k ,\, \cD_l\rcurl = - 2 \partder{}{t_{r(k+l-1)}}$
as in the original unconstrained case \rf{MR-D-n} (modulo $r$). 
\label{proposition:modified-odd}
\end{proposition}

\mark For bosonic \SKPrm models (Eq.\rf{SKP-r-m} with $r,m = even$) there is
no need to modify MR fermionic flows, since in this case 
the term in r.h.s. of \rf{naive-D-n} is absent.

\section*{Berezinian Solutions for the Super-Tau Function}

It was demostrated in \ct{match} that for the general MR-SKP
hierarchy \rf{super-Lax} the \DB (DB) transformations
~${\wti \cL} = \cT \cL \cT^{-1}$ , where $\cT = \chi \cD \chi^{-1}$ with
$\chi$ being a bosonic sEF \rf{super-EF-eqs}
of $\cL$, do {\em not} preserve the fermionic-flow Lax Eqs.\rf{super-Lax-odd}.
Indeed, for the DB-transformed ${\wti \cL}$ to obey the same MR flow 
Eqs.\rf{super-Lax-even}--\rf{super-Lax-odd} as $\cL$, 
the DB-generating ``gauge'' transformation $\cT$ must satisfy:
\be
\partder{}{t_l}\cT \, \cT^{-1} + \(\cT \cL^{2l}_{+} \cT^{-1}\)_{-} = 0
\quad ,\quad 
D_n\cT \, \cT^{-1} - \(\cT \cL^{2n-1}_{+} \cT^{-1}\)_{-} =
- 2 \({\wti \cL}^{2n-1}\)_{-} 
\lab{DB-consist-0}
\ee

The first Eq.\rf{DB-consist-0} is exactly analogous to the purely
bosonic case and implies that $\chi$ must be a sEF \rf{super-EF-eqs} of $\cL$
w.r.t. the even MR-SKP flows. However,the second Eq.\rf{DB-consist-0} does not
have solutions for $\chi$ for the general MR-SKP hierarchy.
In particular, if $\chi$ would be a sEF also w.r.t. fermionic flows
(cf. second Eq.\rf{super-EF-eqs}), then the l.h.s. of second 
Eq.\rf{DB-consist-0} would become zero thereby leading to the contradictory 
relation: $\({\wti \cL}^{2n-1}\)_{-} = 0$.
This makes the standard DB method inapplicable to find solutions of the
unconstrained MR-SKP. 

On the other hand, for constrained fermionic \SKPrm 
hierarchies it can easily be shown (extending the proof given in \ct{match} 
for the simplest \SKPhh case), that {\em auto-}DB transformations ({\sl i.e.}, 
those preserving the constrained form \rf{SKP-r-m} of the initial \SKPrm 
hierarchy) are compatible with the {\em modified} fermionic flows 
\rf{odd-flow-new}--\rf{Psi-j-flow-new}. This latter result guarantees that
any iteration of DB transformations of the initial \SKPrm hierarchy (in
particular, the ``free'' one with $\cL_{({r\o 2},{m\o 2})}^{(0)} = \cD^r$)
will yield new nontrivial solutions for \SKPrm , which obey the same
isospectral flow 
Eqs.\rf{super-Lax-even},\rf{odd-flow-new},\rf{P-j-flow-new}--\rf{Psi-j-flow-new},
{\sl i.e.}, both bosonic {\em and} fermionic, as the initial hierarchy. 

Now, consider auto-DB transformations for arbitrary \SKPrm \rf{SKP-r-m}
(here $L\equiv \cL_{({r\o 2},{m\o 2})} \\
\equiv \cL^{(0)}_{({r\o 2},{m\o 2})}$) :
\br
{\wti L}\! &=& \!\cT_a L \cT_a^{-1} = 
{\wti L}_{+} + \sum_{j=0}^{m-1}{\wti \P}_{{m-j-1}\o 2}\cD^{-1}{\wti \Psi}_{j/2}
\lab{DB-L-r-m} \\
{\wti \P}_a\! &=& \!\cT_a L (\P_a) \quad , \quad 
{\wti \Psi}_a = \P_{{m-2a-1}\o 2}^{-1} , \quad 
{\wti \P}_{{m-j-1}\o 2} = \cT_a (\P_{{m-j-1}\o 2}) \nonu \\
{\wti \Psi}_{j/2}\! &=& \! (-1)^{j+1} {\cT_a^\ast}^{-1} (\Psi_{j/2}) =
(-1)^j \P_a^{-1} \Dth^{-1} (\P_a \Psi_{j\o 2}) \quad {\rm for} \;\;
j \neq m-2a-1
\lab{DB-r-m} 
\er
where $\cT_a =\P_a \cD \P_a^{-1}$ with $a$ being a fixed integer (bosonic)
index. 
Under DB transformations the super-tau function transforms as 
(cf. Eq.(3.4) in \ct{match}) :
\be
{\wti \t} = \P_a \t^{-1}
\lab{DB-s-tau}
\ee

Before proceeding to the iteration of DB-transformations for \SKPrm
hierarchies \rf{DB-L-r-m}--\rf{DB-r-m}, we will introduce some convenient
short-hand notations for Wronskian-type Berezinians:
\br
&&{\rm Ber}_{(k,l)} \lb\vp_0,\ldots ,\vp_{k-1} ;\vp_\h ,\ldots ,\vp_{l-\h}\rb 
\equiv  \nonu \\
&&{\rm Ber} \threemat{\cW_{k,k} \lb \vp_0,\ldots ,\vp_{k-1} \rb}{|}{
\cW_{k,l} \lb \vp_\h ,\ldots ,\vp_{l-\h}\rb}{------------}{|}{------------}{ 
\cW_{l,k} \lb \Dth\vp_0 ,\ldots ,\Dth \vp_{k-1} \rb}{|}{
\cW_{l,l} \lb \Dth\vp_\h ,\ldots ,\Dth\vp_{l-\h}\rb}
\lab{Ber-k-l} 
\er
where $\(\vp_0,\ldots ,\vp_{k-1}\)$ and $\(\vp_\h ,\ldots ,\vp_{l-\h}\)$ are
sets of bosonic (fermionic) superfield functions, and 
where $\cW_{k,l}\lb f_1,\ldots , f_l \rb$ denotes a {\em rectangular}
($k$ rows by $l$ columns) Wronskian matrix:
\be
\cW_{k,l}\lb f_1,\ldots , f_l \rb = 
\left\Vert \pa_x^{\a -1} f_\b \right\Vert  \quad ,\quad
\a =1,\ldots ,k \;\; ,\;\; \b =1,\ldots ,l  
\lab{rect-wronski}
\ee

The derivation of the explicit form of the DB-orbit for the super-tau
function and the (adjoint-)EF's of \SKPrm is based on the following
Proposition:
\begin{proposition}
The iteration of \DB-like transformations on arbitrary
initial superfield functions ($\P$ -- bosonic, $F$ -- fermionic) can be
expressed in a Berezinian form as follows:
\be
\P^{(2n)} \equiv \cT^{(2n-1)}_{\vp_{n-\h}} \cT^{(2n-2)}_{\vp_{n-1}} \ldots
\cT^{(3)}_{\vp_{3/2}} \cT^{(2)}_{\vp_1} \cT^{(1)}_{\vp_\h} \cT^{(0)}_{\vp_0}
(\P ) =  
\lab{Phi-Ber-2n}  
\ee

\vspace{-8.mm}

\br
\({\rm Ber}_{(n,n)} \lb \vp_0,\ldots ,\vp_{n-1},\P ;
\vp_\h ,\ldots ,\vp_{n-\h}\rb\)^{-1}
{\rm Ber}_{(n+1,n)}\lb\vp_0,\ldots ,\vp_{n-1},\P ;\vp_\h ,\ldots ,\vp_{n-\h}\rb
\nonu
\er
\be
F^{(2n+1)} \equiv \cT^{(2n)}_{\vp_n} \cT^{(2n-1)}_{\vp_{n-\h}} \ldots
\cT^{(3)}_{\vp_{3/2}} \cT^{(2)}_{\vp_1} \cT^{(1)}_{\vp_\h} \cT^{(0)}_{\vp_0}
(F) = 
\lab{Phi-Ber-2n+1} 
\ee

\vspace{-7.mm}

\br
{\rm Ber}_{(n+1,n)} \lb \vp_0,\ldots ,\vp_n ;\vp_\h ,\ldots ,\vp_{n-\h}\rb \,
\({\rm Ber}_{(n+1,n+1)}\lb\vp_0,\ldots ,\vp_n ;
\vp_\h ,\ldots ,\vp_{n-\h},F\rb\)^{-1}    \nonu 
\er
where by definition:
\be
\cT^{(j)}_{\vp_{j\o 2}} = \vp_{j\o 2}^{(j)} \cD \frac{1}{\vp_{j\o 2}^{(j)}} 
\quad ,\quad
\vp_{j\o 2}^{(j)} = \cT^{(j-1)}_{\vp_{{j-1}\o 2}} \cT^{(j-2)}_{\vp_{{j\o 2}-1}}
\ldots \cT^{(2)}_{\vp_1} \cT^{(1)}_{\vp_\h} \cT^{(0)}_{\vp_0} (\vp_{j\o 2})
\lab{DB-vp-k}
\ee
\label{proposition:main-ber}
\end{proposition}

Here and in what follows the superscripts in brackets indicate the step of
iteration of DB(-like) transformations.
Note that $F^{(2n+1)}$ \rf{Phi-Ber-2n+1} and $\vp_{k+\h}^{(2k+1)}$ 
\rf{DB-vp-k} are bosonic although the initial $F, \vp_{k+\h}$ are fermionic.

The proof of Prop.\ref{proposition:main-ber} relies on the observation,
that both sides of \rf{Phi-Ber-2n} and \rf{Phi-Ber-2n+1}
define monic super-differential operators acting on $\P$ and $F$,
respectively, which share the same set of kernel elements, namely, the 
superfield functions $\vp_0,\ldots ,\vp_{n-1};\vp_\h ,\ldots ,\vp_{n-\h}$.

Let us consider in more detail the DB-orbit of constrained \SKPrm hierarchy
with $r=1$, {\sl i.e.}, $L \equiv \cL_{(\h,{m\o 2})} = \cD + f_0 +
\sum_{j=0}^{m-1} \P_{{m-1-j}\o 2} \cD^{-1} \Psi_{j\o 2}$
(in the formulas below $m$ indicates the order of the pseudo-differential part
of $L \equiv \cL_{(\h,{m\o 2})}$, the integer $k$ is $0 \leq k \leq m-1$, and 
$l$ is arbitrary non-negative integer) :
\br
\P^{(ml+k)} = \(\cT^{(lm-1+k)}_{\P_{{k-1}\o 2}} \ldots \cT^{(lm)}_{\P_0}\)
\cdots \(\cT^{(m-1)}_{\P_{{m-1}\o 2}} \ldots \cT^{(0)}_{\P_0}\)
\( L^{l+1} (\P_{j\o 2})\)  \nonu \\
{\rm for} \;\; 0 \leq j \leq k-1   \lab{DB-orbit-EF-1} \\
\P^{(ml+k)} = \(\cT^{(lm-1+k)}_{\P_{{k-1}\o 2}} \ldots \cT^{(lm)}_{\P_0}\)
\cdots \(\cT^{(m-1)}_{\P_{{m-1}\o 2}} \ldots \cT^{(0)}_{\P_0}\)
\( L^l (\P_{j\o 2})\)   \nonu \\
{\rm for} \;\; k \leq j \leq m-1   \lab{DB-orbit-EF-2} 
\er
Eqs.\rf{DB-orbit-EF-1}--\rf{DB-orbit-EF-2} indicate that the DB-orbit is
defined by successive iterations of DB-trans\-for\-ma\-tions w.r.t. all 
super-EF's $\P_{j\o 2}$ ($j=0,\ldots , m-1$) present in 
$L \equiv \cL_{(\h,{m\o 2})}$ .
Comparing \rf{DB-orbit-EF-1}--\rf{DB-orbit-EF-2}
with the general formulas \rf{Phi-Ber-2n}--\rf{Phi-Ber-2n+1} 
we easily identify the functions $\vp_k$ and $\vp_{k\o 2}$ appearing in the 
latter with the super-EF's $\P_{j\o 2}$ of $L \equiv \cL_{(\h,{m\o 2})}$ as 
follows:
\be
\vp_{{ml+j}\o 2} = L^l (\P_{j\o 2})
\lab{vp-P}
\ee
Therefore, the explicit expressions for the super-tau functions on the 
DB-orbit \rf{DB-orbit-EF-1}--\rf{DB-orbit-EF-2}, upon using
\rf{DB-s-tau} and \rf{Phi-Ber-2n}--\rf{Phi-Ber-2n+1}, are given by:
\br
\t^{(2n+1)} &=&
{\rm Ber}_{(n+1,n)} \lb\vp_0,\ldots ,\vp_n ;\vp_\h ,\ldots ,\vp_{n-\h}\rb\,
{1\o {\t^{(0)}}}
\lab{DB-s-tau-1} \\
\t^{(2n)} &=& \({\rm Ber}_{(n,n)}\lb\vp_0,\ldots ,\vp_{n-1};\vp_\h ,\ldots,
\vp_{n-\h}\rb\)^{-1} \t^{(0)}
\lab{DB-s-tau-2}
\er
with the substitution \rf{vp-P} for $\vp_{k}, \vp_{k\o 2}$ in the r.h.s. of 
\rf{DB-s-tau-1}--\rf{DB-s-tau-2}.

\section*{Super-Soliton Solutions}

Now, let us provide some explicit examples of Berezinian solutions for the
\SKPrm tau-function \rf{DB-s-tau-1}--\rf{DB-s-tau-2}. We shall
consider the simplest case of constrained \SKPhh hierarchy and take the
initial $\t^{(0)}\! =\! const$, {\sl i.e.}, the initial super-Lax operator being
$L \equiv \cL_{(\h ,\h )} = \cD$. The initial super-EF 
$\P_0 \equiv \P_0^{(0)}$ satisfies according to \rf{P-j-flow-new} :
\br
\partder{}{t_k} \P_0 &=& \pa_x^k \P_0 \quad ,\quad
\cD_n \P_0 = - \Dth^{2n-1} \P_0
\lab{free-sEF-0} \\
\P_0 (t,\th ) &=& 
\int d\l\, \Bigl\lb \vp_B (\l) + 
\Bigl(\th - \sum_{n\geq 1} \l^{n-1} \th_n \Bigr) \vp_F (\l) \Bigr\rb
e^{\sum_{l\geq 1} \l^l (t_l + \th \th_l )}  
\lab{free-sEF}
\er
where $\vp_B (\l),\, \vp_F (\l)$ are arbitrary bosonic (fermionic) 
``spectral'' densities. 

It is easy to show that for \SKPhh case the Berezinian expressions 
\rf{DB-s-tau-1}--\rf{DB-s-tau-2}, together with the
substitution \rf{vp-P}, which now ($m=1,\, j=0$) becomes simply
~$\vp_{l\o 2} = \cD^l \P_0$, reduce to the following ratios of ordinary
Wronskians:
\be
\t^{(2n)} = \frac{W_n \llb \pa_x\P_0,\ldots ,\pa_x^n \P_0\rrb}{
W_n \llb\P_0 ,\ldots ,\pa_x^{(n-1)}\P_0 \rrb }   \quad , \quad
\t^{(2n+1)} = \frac{W_{n+1}\llb\P_0 ,\ldots ,\pa_x^n \P_0 \rrb}{
W_n \llb \pa_x \P_0,\ldots ,\pa_x^n \P_0\rrb}
\lab{tau-match}
\ee
where $\P_0$ is given by \rf{free-sEF}. In particular, choosing for the
bosonic (fermionic) ``spectral'' densities in Eq.\rf{free-sEF} 
~$\vp_B (\l) = \sum_{i=1}^N c_i \d (\l - \l_i )$ ,
$\vp_F (\l) = \sum_{i=1}^N \eps_i \d (\l - \l_i )$ ,
where $c_i,\l_i$ and $\eps_i$ are Grassmann-even and Grassmann-odd
constants, respectively, we have for $\P_0$ :
\be
\P_0 = \sum_{i=1}^N \Bigl\lb c_i + 
\Bigl(\th - \sum_{n\geq 1} \l_i^{n-1} \th_n \Bigr) \Bigr\rb
e^{\sum_{l\geq 1} \l_i^l (t_l + \th \th_l )}  
\lab{free-sEF-00}
\ee
Substituting \rf{free-sEF-00} into \rf{tau-match} we obtain the following
``super-soliton'' solutions for the super-tau function:
\br
&&\t^{(2n+1)} = \frac{\sum_{1 \leq i_1 < \ldots < i_{n+1} \leq N}
{N \choose n+1} {\wti c}_{i_1} \ldots {\wti c}_{i_{n+1}}
E_{i_1} \ldots E_{i_{n+1}} \D_{n+1}^2 (\l_{i_1},\ldots ,\l_{i_{n+1}})}{
\sum_{1 \leq j_1 < \ldots < j_n \leq N}
{N \choose n} {\wti c}_{j_1} \ldots {\wti c}_{j_n} E_{j_1} \ldots E_{j_n} 
\l_{j_1} \ldots \l_{j_n} \D_{n}^2 (\l_{j_1},\ldots ,\l_{j_n})}  \nonu \\
&&\phantom{aaaaaaaaaaaaaaaaaaaa} \lab{tau-super-sol} \\
&&{\wti c}_i \equiv c_i + \Bigl(\th - \sum_{n\geq 1} \l_i^{n-1} \th_n \Bigr)
\quad ,\quad
E_i \equiv e^{\sum_{l\geq 1} \l_i^l (t_l + \th \th_l )}   \nonu \\
&&\D_{n} (\l_{i_1},\ldots ,\l_{i_n}) \equiv 
\det\left\Vert \l_{i_a}^{b-1}\right\Vert_{a,b=1,\ldots ,n}
\lab{super-sol-notation}
\er
\mskp
{\bf Outlook.} There is a number of interesting issues, related to the present 
topic, which deserve further study such as: {\em binary} DB-transformations
and new types of solutions for the super-tau-function; consistent
formulation of supersymmetric two-dimensional Toda lattice and search for 
proper supersymmetric counterparts of random (multi-)matrix models, based on
analogous approach \ct{hungry} in the purely bosonic case.

{\bf Acknowledgements.} The authors gratefully acknowledge support by NSF 
grant {\sl INT-9724747}.

\end{document}